\documentclass[a4paper]{article}
\usepackage[
    pdfauthor={Erica Bertolini, Hyungrok Kim},
    pdftitle={Covariant interacting fractons},
    pdfsubject={hep-th,cond-mat.str-el,math-ph,gr-qc},
    breaklinks=true,
    final
]{hyperref}
\usepackage{amsmath,amssymb,orcidlink,cleveref}
\usepackage[final,spacing]{microtype}
\usepackage[varbb]{newpxmath}
\usepackage[osf]{newpxtext}
\usepackage[russian,ukrainian,greek,english,british]{babel}

\renewcommand\delta\deltaup
\usepackage{CJKutf8}

\title{Covariant interacting fractons}
\author{Erica Bertolini\orcidlink{0000-0003-0773-7526}\footnote{School of Theoretical Physics, Dublin Institute for Advanced Studies, 10 Burlington Road, \textsc{d04~c932}, Dublin, Ireland.\\Scoil na Fisice Teoiriciúla, Institiúid Ard-Léinn Bhaile Átha Cliath, 10 Bóthar Burlington, \textsc{d04~c932}, Baile Átha Cliath, Éire.}\\\href{mailto:ebertolini@stp.dias.ie}{\texttt{ebertolini@stp.dias.ie}}\\Hyungrok Kim \begin{CJK*}{UTF8}{bsmi}(金炯錄)\end{CJK*}\orcidlink{0000-0001-7909-4510}\footnote{Department of Mathematics and Theoretical Physics, University of Hertfordshire, Hatfield, Hertfordshire \textsc{al10~9ab}, United Kingdom.}\\\href{mailto:h.kim2@herts.ac.uk}{\texttt{h.kim2@herts.ac.uk}}}
\begin{document}
\maketitle
\begin{abstract}
We show that there exists a natural analogue of the Yang--Mills equations using the Frölicher--Nijenhuis bracket between vector-valued differential forms. The gauge field is a rank-two tensor, and when one constrains it to be symmetric, then the system exhibits fractonic behaviours. In the linearised limit, the constrained equations of motion reduce to those of the covariant fracton model \cite{Bertolini:2022ijb,Blasi:2022mbl,Bertolini:2024gzx}.
\end{abstract}
\tableofcontents
\section{Introduction and summary}
In recent years, a new kind of quasiparticle has emerged from the physics literature: \emph{fractons} \cite{Vijay:2015mka,Vijay:2016phm,Nandkishore:2018sel,Pretko:2020cko}. Originating in lattice models in the context of spin glasses \cite{Chamon:2004lew} and quantum information \cite{Haah:2011drr,Bravyi:2013ort}, fractons have quickly attracted the interest of a wide variety of physicists from condensed matter to mathematical  physics and have been influential in gauge theories and other quantum field theories \cite{Pretko:2016kxt,Pretko:2016lgv,Yan:2018nco,You:2019bvu,Seiberg:2020bhn,Seiberg:2020wsg,Slagle:2020ugk,Caddeo:2022ibe,Huang:2023zhp,Perez:2023uwt}. The main characteristic of fracton quasiparticles is immobility, which is also the reason for the name. Indeed a fracton is defined as a fraction of a mobile quasiparticle, and in isolation it cannot move at all. Only dipole-like excitations are free to displace, or, in general have fewer constraints on their motion \cite{Pretko:2016kxt,Pretko:2016lgv,Nandkishore:2018sel,Pretko:2020cko}. These additional constraints defines other fracton-related quasiparticles such as \emph{lineons}  and \emph{planons}, which can move in a one- or two-dimensional subspace respectively. The restricted-motion feature, which unites all fracton theories, is shared by many physical systems and models, and it is one of the reasons for which fractons are so popular nowadays. For instance limited mobility, or complete immobility,  can be harnessed for developing quantum memories \cite{Haah:2011drr,Bravyi:2013ort}, or used as a mapping/duality to study topological defects in elastic media \cite{Pretko:2017kvd,Pretko:2019omh,Gromov:2020yoc,Yuan:2019geh,Wang:2019mtt,Grosvenor:2021rrt,Glodkowski:2022xje,Tsaloukidis:2023jmr}. It is also a characteristic found in Carrollian theories, which thus seem to display some fractonic behaviour \cite{Figueroa-OFarrill:2023vbj,Figueroa-OFarrill:2023qty}. The other reason for attracting so much interest is found in the tensorial nature of fractonic theories. In gauge theory \cite{Pretko:2016kxt,Pretko:2016lgv} these models are indeed typically described in terms of a rank-2 symmetric tensor \(A_{ij}(x)\) (with \(i,j\) spatial indices), which transforms under the gauge transformation
\begin{equation}\label{dfract}
    \delta_\epsilon A_{ij}=\partial_i\partial_j\epsilon
\end{equation}
and share strong similarities with the electromagnetic  Maxwell theory, of which they represent higher-rank generalisations. A generalised Gauss law is typically postulated as
\begin{equation}
    \partial_i\partial_jE^{ij}=\rho,
\end{equation}
where \(E^{ij}(x)\) is a symmetric electric tensor field, implying dipole moment conservation through
\begin{equation}
    D^i\coloneqq\int\mathrm d^dx\,x^i\rho=-\int\mathrm d^dx\,\partial_jE^{ij}= \oint\mathrm d^{d-1}x\,(\dotsb),
\end{equation}
that is, the dipole moment cannot change except through a nonzero flux at the boundary.
This encodes the immobility of the fractonic charge \(\rho(x)\) \cite{Nandkishore:2018sel,Pretko:2020cko} since, if a single charge were to move, it would change the total dipole moment of the system. The tensorial nature of the gauge field also hints towards connections with the theory of linearised gravity \cite{Gu:2006vw,Xu:2006pxf,Gu:2009jh,Xu:2010eg,Pretko:2017fbf}, which emerges naturally when the covariant fracton theory is taken into account \cite{Bertolini:2022ijb,Blasi:2022mbl,Bertolini:2023juh,Bertolini:2023sqa}. The covariant extension 
\begin{equation}\label{cov-dfract}
    \delta A_{\mu\nu}=\partial_\mu\partial_\nu\epsilon
\end{equation}
of the fractonic transformation \eqref{dfract} is indeed a particular case of the infinitesimal diffeomorphisms that defines linearised gravity, sometimes called longitudinal diffeomorphisms \cite{Dalmazi:2020xou}. Thus an action  invariant under \eqref{cov-dfract} would naturally carry a linearised gravity term. This covariant formulation \cite{Bertolini:2022ijb} gives rise to the definition of an invariant rank-3 field strength \(F_{\mu\nu\rho}(x)\), through which the Maxwell analogy and fracton phenomenology of \cite{Pretko:2016kxt,Pretko:2016lgv,Prem:2017kxc} is reproduced and expanded from first principles. The covariant theory of fractons \cite{Bertolini:2022ijb} is free (or at least quantum-electrodynamics-like) in the sense that the gauge field \(A_{\mu\nu}(x)\) does not interact directly with itself.
However, the existence of a generalised invariant field strength \(F_{\mu\nu\rho}(x)\) suggests a mathematically natural `nonabelianisation' of the above theory through the use of the Frölicher--Nijenhuis bracket \cite{MR82554,MR132493} (reviewed in \cite[§8]{MR1202431}), which is the focus of this paper.

In an interacting theory of fractons where the gauge field interacts with itself, the gauge field modes themselves are fractonic, so that there may be constraints on asymptotic in- and out-states. This may be interesting from an amplitude-theoretic point of view and may possibly signal new loopholes to Weinberg--Witten-type no-go theorems \cite{Weinberg:1980kq}.

The analysis presented in this paper is also of independent mathematical interest. The Frölicher--Nijenhuis bracket is a fundamental geometric structure, which has recently appeared in the context of integrable models such as self-dual Yang--Mills theory \cite{Muller-Hoissen:2024cva}. In fact, the Frölicher--Nijenhuis bracket generalises to arbitrary Lie algebroids \cite{2014arXiv1412.2533D}, which appear in gauge theory in many contexts \cite{Bojowald:2004wu,Mayer:2009wf,Kotov:2015nuz,Fischer:2020lri,Fischer:2021glc,Fischer:2024vak,Fischer:2024hqy}, and both the fractonic case at hand and ordinary Yang--Mills theory may be seen as special cases (for the tangent Lie algebroid and a Lie algebroid bundle, respectively) of a more general construction associated to general Lie algebroid. We thus see, again, that strong analogies manifest themselves between fractons and Maxwell/Yang--Mills theories.

Our interacting fractonic model is defined on flat space, as gauge invariance breaks on curved space: the field strength \(F\) only transforms tensorially if one assumes that the metric \(g_{\mu\nu}\) is flat and also transforms under diffeomorphism. Thus, to write an action principle one cannot have \(g_{\mu\nu}\) as a background (since it must transform), and the equations of motion for \(g_{\mu\nu}\) must ensure that it remains flat (since otherwise gauge invariance fails). As a consequence, for the scope of the analysis presented here, we only postulate an equation of motion.
For efforts at fractonic behaviours on curved spaces, see \cite{Slagle:2018kqf,Afxonidis:2023pdq,Rovere:2024nwc}. 
We also do not discuss issues regarding the classical or quantum stability of our model.

\paragraph{Acknowledgements}
Hyungrok Kim thanks Simon-Raphael Fischer\orcidlink{0000-0002-5859-2825} and Fridrich Valach~III\orcidlink{0000-0003-0020-1999} for helpful discussion.

\section{Mathematical background}
In the following, we will need to make use of the Frölicher--Nijenhuis bracket on vector-valued differential forms and the language of twisting, which we review briefly.

\subsection{Frölicher--Nijenhuis bracket}
Let \(M\) be a smooth manifold. The graded vector space of vector-valued differential forms \(\Omega^\bullet(M;\mathrm TM)=\bigoplus_{i=0}^d\Omega^i(M)\) becomes a graded Lie algebra with respect to the \emph{Frölicher--Nijenhuis bracket} \cite{MR82554,MR132493} (reviewed in \cite[§8]{MR1202431}):
\begin{multline}\label{eq:Froelicher-Nijenhuis}
[\phi \otimes X,\psi \otimes Y]=(\phi \wedge \psi)\otimes\mathcal L_XY+(\phi \wedge\mathcal L_X\psi) \otimes Y-(\mathcal L_Y\phi \wedge \psi) \otimes X\\+(-1)^p(\mathrm d\phi \wedge i_X\psi)\otimes Y+(-1)^p(i_Y\phi\wedge\mathrm d\psi)\otimes X
\end{multline}
for vector fields \(X,Y\in\Gamma(\mathrm TM)\) (where \(\Gamma(-)\) denotes the space of sections of a vector bundle) and homogeneous differential forms \(\phi\in\Omega^p(M)\), \(\psi\in\Omega^q(M)\), where \(\mathcal L_X(-)\) is the Lie derivative of a tensor field along a vector field, and \(i_X(-)\) is the interior derivative of a differential form along a vector field.

In particular, between two (1,1)-tensors \(K^\mu_\nu\) and \(L^\mu_\nu\), we have
\begin{equation}[K,L]_{\mu\nu}^\rho=
2K_{[\mu|}^\sigma\partial_\sigma L_{|\nu]}^\rho
+
2L^\sigma_{[\mu|}\partial_\sigma K^\rho_{|\nu]}
-2K^\rho_\sigma\partial_{[\mu}L^\sigma_{\nu]}
-2L^\rho_\sigma \partial_{[\mu}K^\sigma_{\nu]}.\end{equation}
where antisymmetrisations are normalised.

Notice that, when one of the arguments is a (1,0)-tensor (i.e.~a vector field), it reduces to the usual Lie derivative:
\begin{equation}
    [X,-]=\mathcal L_X\qquad(X\in\Gamma(\mathrm TM)).
\end{equation}

Suppose that \(M\) is equipped with a Riemannian metric \(g\) whose Riemann curvature vanishes. Then, using the induced Levi-Civita connection \(\nabla\),
we may define the covariant exterior derivative
\begin{equation}
    \mathrm d^\nabla\colon \Omega^\bullet(M;\mathrm TM)
    \to\Omega^{\bullet+1}(M;\mathrm TM)
\end{equation}
that squares to zero, and then \(\Omega^\bullet(M;\mathrm TM)\) forms a differential graded Lie algebra.
Note that, when the curvature of \(g\) does not vanish, then \(\mathrm d^\nabla\) need not square to zero.

\subsection{Twisting}\label{ssec:twist}
A curved\footnote{The nomenclature comes from an analogy with the fact that, when one considers e.g.\ differential forms valued in a vector bundle with connection, the (covariant) exterior derivative operator does not in general square to zero anymore but rather to a curvature-dependent term.} differential graded Lie algebra \((\mathfrak g,r,d,[-,-])\) is a \(\mathbb Z\)-graded Lie algebra \((\mathfrak g,[-,-])\) together with a linear map
\begin{equation}
    d\colon\mathfrak g\to\mathfrak g
\end{equation}
of degree one and an element \(r\in \mathfrak g\) of degree two, called the \emph{curvature}, such that \(d\) is a graded derivation with respect to the Lie bracket \([-,-]\) and
\begin{equation}
    d(d(x)) = [r,x]
\end{equation}
for any \(x\in\mathfrak g\).
This is the special case of the notion of a curved \(L_\infty\)-algebra \cite{Kraft:2022efy,2018arXiv181002941D,Dotsenko:2022bbv} \((\mathfrak g,\mu_0,\mu_1,\mu_2,\dotsc)\), which is a graded vector space \(\mathfrak g\) equipped with totally graded-antisymmetric multilinear maps \(\mu_i\colon\mathfrak g^{\otimes i}\to\mathfrak g\) that satisfy the Jacobi identity up to homotopy. Then a curved differential graded Lie algebra is the same as a curved \(L_\infty\)-algebra in which \(\mu_i=0\) except for \(i\in\{0,1,2\}\); then \(\mu_0\), \(\mu_1\), and \(\mu_2\) correspond to \(r\), \(d\), and \([-,-]\), respectively.

Given a differential graded Lie algebra \((\mathfrak g,d_{\mathfrak g},[-,-]_{\mathfrak g})\) and an element \(Q\in \mathfrak g^1\) of degree one, then the \emph{twist}\footnote{This is a special case of the twisting of a curved \(L_\infty\)-algebra \cite{Kraft:2022efy,2018arXiv181002941D,Dotsenko:2022bbv}.} of \(\mathfrak g\) by \(Q\) is the curved differential graded Lie algebra \(\mathfrak g_Q\coloneqq(\mathfrak g,dQ+\frac12[Q,Q]_{\mathfrak g},d_{\mathfrak g}+[Q,-]_{\mathfrak g},[-,-]_{\mathfrak g})\). When \([Q,Q]_{\mathfrak g}=0\), then this is a differential graded Lie algebra.

\section{Motivation and idea}\label{sec:motivation}
\subsection{A review of Yang--Mills theory}
Linearised Yang--Mills theory (that is, a direct sum of copies of Maxwell theory) admits a natural nonabelianisation in the form of Yang--Mills theory. Let us recall how this works. The field strength in linearised Yang--Mills theory is
\begin{equation}
    F^a_{\mu\nu} = 2\partial_{[\mu} A^a_{\nu]}.
\end{equation}
In the non-Abelian theory, this is modified to
\begin{equation}
    F^a_{\mu\nu} = 2\partial_{[\mu} A^a_{\nu]}+f^a{}_{bc}A^b_\mu A^c_\nu.
\end{equation}
It is convenient to use the notation of Lie-algebra-valued differential forms, in terms of which we have
\begin{align}
    F&=\mathrm dA+\frac12[A,A].
\end{align}
That is, the field strength is fixed by the structure of a differential graded Lie algebra on the space of \(\mathfrak g\)-valued differential forms \(\Omega^\bullet(M)\otimes\mathfrak g=\bigoplus_{i=0}^{\dim M}\Omega^i(M)\otimes\mathfrak g\), where \(\mathfrak g\) is the colour Lie algebra, \(M\) is spacetime, and \(^\bullet\) is a placeholder for the form degree. Furthermore, this fixes the structure of gauge transformations and Bianchi identities:
\begin{align}\label{eq:YM_gauge_transformation}
    \delta_\alpha A &= \mathrm d_A\alpha &
    \mathrm d_A F&= 0 &
    \mathrm d_A &\coloneqq \mathrm d+\frac12[A,-]&
    \delta_\alpha F &= [\alpha,F].
\end{align}
The procedure of replacing \(\mathrm d\) with \(\mathrm d_A\) goes by the name of \emph{twisting} \cite{Kraft:2022efy,2018arXiv181002941D,Dotsenko:2022bbv} as discussed in \cref{ssec:twist}. After this, we no longer have a differential graded Lie algebra in the usual sense since
\begin{equation}
    \mathrm d_A^2 = [F,-]\ne0,
\end{equation}
but we speak of a \emph{curved} differential graded Lie algebra. Given this, the equation of motion for the theory is fixed to be
\begin{equation}
    \mathrm d_A\star F = 0.
\end{equation}

\subsection{Nonabelianising the covariant fracton model: the idea}
The covariant fracton model \cite{Bertolini:2022ijb} is a free theory whose  fundamental field is a symmetric tensor \(A_{\mu\nu}(x)\).
An invariant field strength with one derivative can be defined, which is of the form \cite[eq.\ 7.2.16]{Bertolini:2024gzx}
\begin{equation}
    F_{\mu\nu\rho}=a_1\partial_\mu A_{\nu\rho}+a_2\partial_\rho A_{\mu\nu}-(a_1+a_2)\partial_\nu A_{\mu\rho}
\end{equation}
for some suitable parameters \(a_1,a_2\in\mathbb R\).
For any value of \(a_1,a_2\), the field strength \(F_{\mu\nu\rho}(x)\) satisfies a Bianchi identity \cite[p.~83]{Bertolini:2024gzx}
\begin{equation}
    0=
    \partial_\mu(F_{\beta\nu\rho}-F_{\beta\rho\nu})
    +
    \partial_\nu(F_{\beta\rho\mu}-F_{\beta\mu\rho})
    +
    \partial_\rho(F_{\beta\mu\nu}-F_{\beta\nu\mu})
    =6\partial_{[\mu|}F_{\beta|\nu\rho]}.
\end{equation}
This has the form of an exterior derivative, except that the index \(\beta\) does not participate in the antisymmetrisation. Thus, it is natural to regard \(F\) as a \(\mathrm T^*M\)-valued two-form, similar to how the Yang--Mills field strength is a Lie-algebra-valued two-form. This then means that \(A_{\mu\nu}(x)\) should also be regarded as a \(\mathrm T^*M\)-valued one-form.

There are however three problems that arise in this case, which are related.
\begin{enumerate}
\item There is no obvious Lie bracket for \(\mathrm T^*M\)-valued differential forms (unlike Lie-algebra-valued differential forms).
\item A \(\mathrm T^*M\)-valued one-form will \emph{not} generally be antisymmetric between its two indices.
\item The gauge parameter should naturally be a \(\mathrm T^*M\)-valued zero-form, i.e.\ an ordinary one-form, which is bigger than the scalar field gauge parameter of the covariant fracton model.
\end{enumerate}
We resolve these interrelated problems as follows.
\begin{enumerate}
\item Unlike \(\mathrm T^*M\)-valued forms, there \emph{does} exist a natural Lie bracket on \(\mathrm TM\)-valued forms: the Frölicher--Nijenhuis bracket \cite{MR82554,MR132493} (reviewed in \cite[§8]{MR1202431}). Thus, we work with \(\mathrm TM\)-valued forms, and initially ignore the symmetry property of the \(\mathrm TM\)-valued one-form gauge field \(A^\mu{}_\nu(x)\). Therefore, \(\Omega^\bullet(M;\mathrm TM)\) is a graded Lie algebra. For this to be a \emph{differential} graded Lie algebra, we fix a flat connection on \(\mathrm TM\).
\item Having formulated this theory, then we will impose the symmetry requirement with respect to a pseudo-Riemannian metric:
\begin{equation}\label{eq:symmetry_condition}
    g_{\mu\nu}A^\nu{}_\rho=g_{\rho\nu}A^\nu{}_\mu.
\end{equation}
\item The constraint \eqref{eq:symmetry_condition} will then naturally reduce the gauge symmetry from \(\Omega^0(M;\mathrm TM)\) to \(\Omega^0(M)\), i.e.\ it will require the gauge parameter to be a scalar as for the covariant fracton theory.
\end{enumerate}

\section{Covariant interacting fractonic gauge theory}
As mentioned in \cref{sec:motivation}, we first construct a nonlinear equation of motion for a (1,1)-tensor field \(A^\mu{}_\nu\) in \cref{ssec:non-symmetric}. Then we constrain it to be symmetric in \cref{ssec:symmetric}.

\subsection{Non-symmetric theory}\label{ssec:non-symmetric}
Let \(M\) be a smooth manifold equipped with a flat Riemannian metric \(g_{\mu\nu}\) (such as Minkowski space). Then \((\Omega^\bullet(M;\mathrm TM),\mathrm d^\nabla,[-,-])\) is a differential graded Lie algebra, where \([-,-]\) is the Frölicher--Nijenhuis bracket \eqref{eq:Froelicher-Nijenhuis}.

Consider a (1,1)-tensor
\begin{equation}
    A^\mu{}_\nu\in\Omega^1(M;\mathrm TM).
\end{equation}
Then we may twist, as discussed in \cref{ssec:twist}, to obtain the curved differential graded Lie algebra
\begin{equation}
    (\Omega^\bullet(M;\mathrm TM),F_A,\mathrm d^\nabla_A,[-\wedge-])
\end{equation}
where
\begin{equation}
    \mathrm d^\nabla_A \coloneqq \mathrm d^\nabla + [A,-]
\end{equation}
and
\begin{equation}
    F_A=\mathrm d^\nabla A+\frac12[A,A]
\end{equation}
is the curvature.\footnote{If one wishes, one can work with a rescaled version of the Frölicher--Nijenhuis bracket, where the right-hand side of \eqref{eq:Froelicher-Nijenhuis} comes with an additional factor of \(g\); then then the \(g\to0\) limit corresponds to the free theory.} In particular, we have
\begin{equation}
    (\mathrm d^\nabla_A)^2 = [F_A,-].
\end{equation}

We postulate the infinitesimal gauge symmetry
\begin{equation}
    \delta_\epsilon A=\mathrm d_A\epsilon
\end{equation}
for a vector gauge parameter \(\epsilon\in\Omega^0(M;\mathrm TM)=\Gamma(\mathrm TM)\) which, in explicit component notation, is
\begin{equation}\label{eq:gauge_transformation_covariant_explicit}
    (\delta_\epsilon A)^\mu{}_\nu=\nabla_\nu \epsilon^\mu
    -\mathcal L_\epsilon A^\nu
    =\partial_\nu \epsilon^\mu
    -\epsilon^\rho\partial_\rho A^\mu{}_\nu
    +\partial_\rho\epsilon^\mu A^\rho{}_\nu
    -\partial_\nu\epsilon^\rho A^\mu{}_\rho,
\end{equation}
and define the field strength
\begin{equation}
    F_A \coloneqq \mathrm dA+\frac12[A,A] \in \Omega^2(X;E).
\end{equation}
Explicitly,
\begin{equation}
    F^\rho{}_{\mu\nu}=\nabla_\mu A^\rho{}_\nu - \nabla_\nu A^\rho{}_\mu + \mathcal O(A^2),
\end{equation}
where \(\nabla_\mu\) is the (Riemannian) covariant derivative of a tensor field. This is, to linear order, similar to the field strength in \cite[(7.2.16)]{Bertolini:2024gzx} with \((a_1,a_2)=(1,0)\), which however depends on a symmetric rank-2 tensor, while here \(A^\mu{}_\rho\) is an arbitrary rank (1,1) tensor. We shall discuss the symmetric case in \cref{ssec:symmetric}.

Under a gauge transformation, the field strength \(F\) then transforms covariantly:
\begin{equation}
    \delta_\epsilon F = -\mathcal L_\epsilon F.
\end{equation}
The fact that it is not invariant reminds us of the field strength in Yang--Mills theory \eqref{eq:YM_gauge_transformation}. If we interpret the gauge parameter \(\epsilon^\mu\) as an infinitesimal diffeomorphism, then this implies that \(F^\mu{}_{\nu\rho}\) transforms tensorially.

Now, we may postulate the equation of motion
\begin{equation}
    \nabla^\nu F^\mu{}_{\nu\rho} = 0.
\end{equation}
This is a diffeomorphism-invariant equation as long as we also transform \(g_{\mu\nu}\) under diffeomorphisms, i.e.
\begin{equation}
    \delta_\epsilon g_{\mu\nu}\coloneqq
    -\mathcal L_\epsilon g_{\mu\nu}
    =-\nabla_\mu \epsilon_\nu-\nabla_\nu\epsilon_\mu.
\end{equation}

\subsection{Symmetric theory}\label{ssec:symmetric}
To make contact with the covariant fracton model, we now constrain \(A^\mu{}_\nu\) to be symmetric. That is, we impose the following constraint:
\begin{equation}
    g_{\mu\nu}A^\nu{}_\rho = g_{\rho\nu}A^\nu{}_\mu.
\end{equation}
This constraint is not gauge-invariant for arbitrary \(\epsilon^\mu\) since the right-hand side of \eqref{eq:gauge_transformation_covariant_explicit} need not be symmetric. However, it \emph{is} gauge-invariant if we restrict to `diffeomorphisms' of the form
\begin{equation}
    \epsilon^\mu = \partial^\mu\lambda
\end{equation}
(known as the longitudinal diffeomorphisms \cite{Dalmazi:2020xou}) for some smooth function \(\lambda\in\mathcal C^\infty(M)\), so that the resulting gauge transformation is
\begin{equation}
    \delta_\epsilon A_{\mu\nu}=\partial_\mu\partial_\nu\lambda+\dotsb,
\end{equation}
which to linearised order agrees with the covariant fracton gauge transformation \eqref{cov-dfract}.
Now, we have the symmetrised equation of motion
\begin{equation}
    \nabla^\mu F_{(\rho|\mu|\nu)} = 0.
\end{equation}
The linearised equation of motion (with \(g_{\mu\nu}=\eta_{\mu\nu}\) the Minkowski metric) is
\begin{equation}
    0 = \partial^\mu(\partial_\mu A_{\rho\nu}-\partial_\nu A_{\rho\mu}) + \partial^\mu(\partial_\mu A_{\nu\rho}-\partial_\rho A_{\nu\mu})
    = 2\partial^2A_{\rho\nu}-\partial_\nu\partial^\mu A_{\rho\mu}-\partial^\mu\partial_\rho A_{\nu\mu},
\end{equation}
which is the equation of motion found in the covariant fracton theory of fractons \cite{Bertolini:2022ijb,Blasi:2022mbl,Bertolini:2024gzx}.

\bibliographystyle{unsrturl}
\bibliography{biblio,biblio2}
\end{document}